\documentclass[%
reprint,
 amsmath,amssymb,
 aps,
pre,
]{revtex4-1}

\usepackage{dcolumn}
\usepackage{bm}
\usepackage{graphicx}
\usepackage{subfigure}
\usepackage{color}
\usepackage{epsfig}
\usepackage{amssymb}
\usepackage{latexsym}

\newtheorem{prop}{Proposition}[section]

\newtheorem{cor}{Corollary}

\newtheorem{lm}{Lemma}

\newtheorem{thm}{Theorem}

\newcommand{\bthm}{\begin{thm}}
\newcommand{\ethm}{\end{thm}}

\newcommand{\bcor}{\begin{cor}}
\newcommand{\ecor}{\end{cor}}
\newcommand{\bprop}{\begin{prop}}
\newcommand{\eprop}{\end{prop}}
\newcommand{\blm}{\begin{lm}}
\newcommand{\elm}{\end{lm}}
\newcommand{\beq}{\begin{equation}}
\newcommand{\eeq}{\end{equation}}
\newcommand{\ber}{\begin{eqnarray}}
\newcommand{\eer}{\end{eqnarray}}

\newenvironment{proof1}{\begin{trivlist}\item[]{\bf Proof:\hspace{2mm}}}{\hfill$\blackbox$\end{trivlist}}

\newcommand{\blackbox}{\vrule height7pt width5pt depth1pt}

\newcommand{\bit}{\begin{itemize}}
\newcommand{\eit}{\end{itemize}}
\newcommand{\ben}{\begin{enumerate}}
\newcommand{\een}{\end{enumerate}}
\newcommand{\bdesc}{\begin{description}}
\newcommand{\edesc}{\end{description}}
\newcommand{\beqarrn}{\begin{eqnarray*}}
\newcommand{\eeqarrn}{\end{eqnarray*}}
\newenvironment{proofof}[1]{\begin{trivlist}\item[]{\bf Proof of #1:\hspace{2mm}
}}{\hfill\blackbox\end{trivlist}}
\newcommand{\bproofof}{\begin{proofof}}
\newcommand{\eproofof}{\end{proofof}}
\newenvironment{rem}{\begin{trivlist}\item[]{\bf
Remark:}\hspace{4mm}}{\end{trivlist}}
\newcommand{\brem}{\begin{rem}}
\newcommand{\erem}{\end{rem}}
\newenvironment{rems}{\begin{trivlist}\item[]{\bf
Remarks}\begin{itemize}}{\end{itemize}\end{trivlist}}
\newcommand{\brems}{\begin{rems}}
\newcommand{\erems}{\end{rems}}
\newtheorem{fact}{Fact}
\newcommand{\bfact}{\begin{fact}}
\newcommand{\efact}{\end{fact}}
\newtheorem{examp}{Example}
\newcommand{\bexamp}{\begin{examp}\rm}
\newcommand{\eexamp}{\end{examp}}
\newtheorem{defn}{Definition}
\newcommand{\bdefn}{\begin{defn}\rm}
\newcommand{\edefn}{\end{defn}}

\newtheorem{prob}{Problem}
\newcommand{\bprob}{\begin{prob}}
\newcommand{\eprob}{\end{prob}}

\newcommand{\bvtm}{\begin{verbatim}}
\newcommand{\bfig}{\begin{figure}}
\newcommand{\efig}{\end{figure}}
\newcommand{\bcen}{\begin{center}}
\newcommand{\ecen}{\end{center}}

\long\def\comment#1{}

\def \n2{{N_0 \over 2}}

\newcommand{\bP}[1]{{\mathbb{P}}\left[{#1}\right]}
\newcommand{\bE}[1]{{\mathbb{E}}\left[{#1}\right]}

\newcommand{\1}[1]{{\bf 1}\left[#1\right]}
\usepackage{url}

\def \h5{\hspace{0.5in}}

\graphicspath{{Figures/}}

\begin{document}

\preprint{APS/123-QED}

\title{Robustness of Flow Networks against Cascading Failures under Partial Load Redistribution}

\author{Omur Ozel, Bruno Sinopoli, Osman Ya\u{g}an}
%
\affiliation{%
 Department of ECE, 
Carnegie Mellon University, Pittsburgh, PA 15213 USA}%




%

\date{\today}

\begin{abstract}
 We study the robustness of flow networks against cascading failures under a {\em partial} load redistribution model. In particular, we consider a flow network of $N$ lines with initial loads $L_1, \ldots, L_N$ and free-spaces (i.e., redundant space) $S_1, \ldots, S_N$ that are independent and identically distributed with joint distribution $P_{LS}(x,y)=\bP{L \leq x, S \leq y}$. The {\em capacity} $C_i$ is the maximum load allowed on line $i$, and is  given by $C_i=L_i + S_i$.
When a line fails due to overloading, it is removed from the system and $(1-\varepsilon)$-fraction of the load it was carrying is redistributed {\em equally} among all remaining lines in the system. The rest (i.e., $\varepsilon$-fraction) of the load is assumed to be {\em lost} or {\em absorbed}, e.g., due to advanced circuitry disconnecting overloaded power lines or an {\em inter-connected} network or material absorbing a fraction of the flow from overloaded lines. We analyze the robustness of this flow network against {\em random} attacks that remove a $p$-{\em fraction} of the lines. Our contributions include (i) deriving the {\em final} fraction of alive lines $n_{\infty}(p,\varepsilon)$ for all $p, \varepsilon \in (0,1)$ and confirming the results via extensive simulations; (ii) showing that partial redistribution might lead to (depending on the parameter $0<\varepsilon \leq 1$) the order of transition at the critical attack size $p^{\star}$ changing from first to second-order; and (iii) proving analytically that flow networks achieve maximum robustness (quantified by the area $\int_{0}^{1} n_{\infty}(p,\varepsilon) \mathrm{d}p$) when all lines have the same free-space regardless of their initial load. The optimality of equal free-space allocation is also confirmed on real-world data from the UK National Power Grid.

\begin{description}
\item[PACS numbers]{64.60.Ht, 62.20.M-, 89.75.-k, 02.50.-r}


\end{description}
\end{abstract}
\maketitle

\section{Introduction}

 Flow network abstractions have been extensively used to analyze complex  phenomena occurring in power line networks, financial networks, transportation networks and biological ecosystems. Current trends in  technological development such as online social media, cyber-physical systems and the internet of things have enabled a plethora of new applications involving dynamical interactions over and within flow networks. In this respect, current research on networks covers phenomena such as information dissemination and influence propagation \cite{ TangYuanMaoLiChenDai,wu2017influence,YaganPRE, zhuang2016information,yagan2013conjoining}, percolation \cite{parshani2010interdependent,son2012percolation,min2014network,lee2014threshold,wu2014multiple,radicchi2015percolation} and robustness \cite{yagan2012optimal,HuangGaoBuldyrevHavlinStanley,li2012cascading,buldyrev2010catastrophic,gao2011robustness,brummitt2012suppressing}.

This paper focuses on the robustness of flow networks. In particular, we are interested in understanding the how the random failure of a fraction lines might trigger failures of other lines leading to what has been identified in the literature as {\em cascading failures}. Real life phenomena such as  blackouts in power networks \cite{dobson2007complex} and  crises in financial networks \cite{elliott2014financial} occur as a consequence of  failures that follow one after the other. Consequently, understanding the possible mechanisms of such failures is of paramount importance; e.g., see \cite{zhang2013understanding}. The robustness of flow network structures against cascading failures is an active research topic and has received recent interest from many researchers \cite{yagan2012optimal,HuangGaoBuldyrevHavlinStanley, buldyrev2010catastrophic}. For example, \cite{PahwaScoglioScala, Daniels1945, AndersenSornetteKwan,motter2017,motterPRL} consider power networks where the failure mechanism is the equal redistribution of load upon the failure of a power line. We consider a similar phenomenon in flow networks. In particular, we build our analysis upon the well known fiber bundle model \cite{yagan_fiber_bundle}. Fiber bundle models have been used in a wide range of applications including fatigue \cite{curtin1993tough}, failure of composite materials \cite{kun2007fatigue} and landslides \cite{cohen2009fiber}.

Flow networks typically include built-in systems to counter cascade formations and alleviate the spread of the adverse effect of failures over other lines in the network \cite{brummitt2012suppressing}. In this paper, our objective is to obtain a unified understanding of cascading failures for networks with a specific mechanism to counter such failures. In particular, we study the robustness under {\em partial} load redistribution in a {\em democratic fiber bundle-like} model. Our problem setting is as follows: We consider $N$ lines whose initial loads $L_1, \ldots, L_N$ and free-spaces $S_1, \ldots, S_N$ have joint distribution $P_{LS}(x,y)=\bP{L \leq x, S \leq y}$ and are independent and identically distributed along lines. The maximum flow allowed on a line $i$ defines its {\em capacity}, and is given by $C_i=L_i+S_i$. When a line fails due to overloading, it is removed from the system and $(1-\varepsilon)$-fraction of the load it was carrying (at the moment of failing) gets redistributed {\em equally} among all remaining lines in the system; hence we refer to this as the {\em partial} load redistribution model. The rest (i.e., $\varepsilon$-fraction) of the load is assumed to be {\em lost} or {\em absorbed}, e.g., due to advanced circuitry disconnecting overloaded power lines or an {\em inter-connected} network/material absorbing a fraction of the flow from overloaded lines. Throughout the paper, we refer to $\varepsilon$ as the loss or absorption factor, interchangeably.  

We study the robustness of the flow network described above against {\em random}  attacks or failures. In particular, we develop a complete analytic framework that unravels the dynamics of cascading failures initiated by a random  attack removing a $p$-{\em fraction} of the lines. This is done by computing recursively the fraction of surviving lines and the additional load distributed to each surviving line at every step of the cascade. We show how these recursive relations can be utilized to derive the {\em final} fraction $n_{\infty}(p,\varepsilon)$ of alive lines, under any attack size $0 < p < 1$ and any absorption factor $\varepsilon \in (0,1)$. These results are also confirmed through an extensive numerical study.

An interesting observation from our results concerns the {\em order} of phase transition that the system exhibits at the {\em critical} attack size $p^{\star}$; this is the smallest attack size $p$ that collapses the entire system. It is widely known  \cite{AndersenSornetteKwan,yagan_fiber_bundle} that flow networks  exhibit only first-order phase transitions; this phenomenon is referred to as the {\em abrupt rupture} in the context of fiber bundles. However, we show in this paper that under the  {\em partial} redistribution model, 
the order of the phase transition might be first or second order depending on the absorption factor $\varepsilon$. In particular, we demonstrate that the order of the phase transition changes from first to second-order after $\varepsilon$ exceeds a certain {\em tricritical} point. This is reminiscent of the behavior observed in percolation studies of inter-connected networks \cite{parshani2010interdependent}, where reducing the coupling strength between two inter-connected networks might lead to a second-order transition \footnote{For the model studied in this paper, the effect of increasing $\varepsilon$ can be thought as a reduction in the dependency (or, coupling) among the $N$ lines forming the flow network.}.


Utilizing the analytic framework we developed, we then seek to answer a very fundamental question concerning flow networks: Under given constraints on the total load and capacity available to $N$ lines, how should these be allocated so that the resulting flow network has the maximum possible robustness against random attacks. This question is particularly challenging under the {\em partial} flow redistribution model owing to the fact that the system's robustness may not be fully characterized by the critical attack size $p^{\star}$. Simply put, it is possible for an allocation to lead to a {\em large} $p^{\star}$ but a very poor performance (in terms of final system size $n_{\infty}(p,\varepsilon)$) for even small attack sizes $p$. In order to meaningfully asses the robustness of a flow network under attacks of all possible scale, we thus consider the robustness metric proposed in \cite{scheider2011} given as
\[
\mathcal{R}(\epsilon) := \int_{0}^{1} n_{\infty}(p,\varepsilon) \mathrm{d}p
\]
It can be seen that $\mathcal{R}(\epsilon)$ measures the
{\em area} that falls under the curve $n_{\infty}(p,\varepsilon)$ as $p$ varies from zero to one. More importantly, $\mathcal{R}(\epsilon)$ gives the {\em mean} fraction of lines that will survive an attack whose size $p$ is random over $(0,1)$. 


Although it is often very difficult (if not impossible) to analytically prove optimality results for any network property, here we 
are able to give a complete mathematical proof showing how the robustness metric $\mathcal{R}(\epsilon)$ can be maximized. 
In particular, we show that for a given total initial load and capacity to be allocated to $N$ lines, robustness $\mathcal{R}(\epsilon)$ is maximized when all lines have the same free-space, i.e., capacity minus initial load, regardless of how the initial loads are allocated. This result is shown to hold under any absorption factor $0 \leq \epsilon \leq 1$.  

The optimality of equal free-space allocation 
(in the sense of maximizing robustness)
in flow networks is interesting for several reasons. In particular, it shows that the widely used \cite{MotterLai,WangChen,Mirzasoleiman,CrucittiLatora} assumption that line capacity is a fixed proportion of its load (e.g., by setting $C_i = (1+\alpha) L_i$ with $\alpha$ defining the {\em tolerance factor} used for all $N$ lines) leads to sub-optimal robustness. In fact, our result shows that in order to maximize robustness, lines with higher initial load should be given a smaller tolerance factor (i.e., free-space divided by load). Moreover, our results show that although a flow network's robustness might be improved by increasing its capability to {\em absorb} failed load (which often will require a more expensive or complicated design), the system can be made optimally robust by allocating all lines the same free-space irrespective of the absorbing capability.

We test the above result under a new cascading failure model for flow networks that combine local and global redistribution approaches; e.g., a portion of the failed load is redistributed in the local neighborhood according to network topology while the rest is redistributed globally. For this model, we provide simulation results that suggest that the mean-field redistribution model analyzed here captures the qualitative behavior of system robustness well.
In particular, we observe that uniformly allocating free-spaces promises to be optimal also in this more general setting where (part of the) failed load gets redistributed locally according to a network topology.

Finally, we test our optimality result on real-world systems. In particular, we run numerical experiments using the UK National Grid data available in \cite{link}. These experiments confirm our optimality result and suggest that some real-world systems do not exhibit optimal robustness against random failures. In fact, their robustness can be improved significantly by re-allocating the total available capacity in such a way as to ensure that every line has the same free-space. We believe that our results provide interesting insights into the dynamics of cascading failures in flow networks and call for a careful examination of the capacity allocations in existing real-world systems.



The rest of the paper is organized as follows. In Section \ref{sec:Model}, we explain the system model in detail 
and present the problem definition. In Section \ref{sec:Results}, we obtain an iterative dynamical relation for the extra load per alive line at every stage of the cascading failure process, under general load and free-space distributions. Section \ref{sec:Simu} is devoted to numerical results that confirm the main findings of the paper for systems of finite size. In Section \ref{sec:Optimal}, we present a complete analytic framework to establish the optimal distribution
of load and free-space (when mean values of both are fixed)
that leads to maximum robustness. These findings are confirmed via extensive simulations in Section \ref{subsec:real-data} using  synthetic data (with various commonly used distributions utilized to generate load and free-space values) as well as real-world data from the UK National Power Grid. The paper is concluded in Section \ref{sec:Conclusion}.

\section{Model and Problem Definition}
\label{sec:Model}

{\bf Partial load-redistribution model.} We consider a network with $N$ lines $\mathcal{L}_1, \ldots, \mathcal{L}_N$ with initial loads $L_1, \ldots, L_N$. The {\em capacity} $C_i$  of a line $\mathcal{L}_i$ defines the maximum power flow that can be carried by it, and is expressed as  
\begin{equation}
C_i= L_i + S_i, \qquad i=1,\ldots, N,
\label{eq:capacity}
\end{equation}
where $S_i$ denotes the {\em free-space} that is assigned to line $\mathcal{L}_i$. Alternatively, the capacity of a line  $\mathcal{L}_i$ can be defined as a factor of its initial load, i.e.,
\begin{equation}
C_i = (1+\alpha_i) L_i
\label{eq:capacity_with_tolerance}
\end{equation}
with $\alpha_i>0$ denoting its {\em tolerance} factor. Accordingly, the free-space $S_i$ is given in terms of the initial load $L_i$ as $S_i = \alpha_i L_i$. Most existing works assume a {\em fixed} tolerance factor for all lines in the system, i.e., $\alpha_i=\alpha$ for all $i$; e.g., see \cite{MotterLai,WangChen,Mirzasoleiman,CrucittiLatora}.

The main assumption of our model is that when a line fails due to {\em overloading}, i.e., due to its load exceeding its capacity, it is removed from the system and $(1-\varepsilon)$-fraction of the load it was carrying (at the moment of failing) gets redistributed {\em equally} among all remaining lines in the system; hence we refer to this as the {\em partial} load redistribution model. The rest (i.e., $\varepsilon$-fraction) of its load is assumed to be {\em lost} or {\em absorbed}.

The partial load redistribution model is motivated by several real-world scenarios. For instance, most real-world power systems are equipped with advanced protection circuits that immediately disconnect the overloaded lines from the rest of the grid \cite{report,shedding}; the parameter $\varepsilon$ would then represent the fraction of lines protected by such advanced circuitry. Alternatively, we can think of a flow network (resp.~a bundle of fibers) that is {\em inter-connected} with another network (resp.~material) that can absorb a fraction of the 
flow from overloaded lines.

Throughout we assume that the pairs  $(L_i, S_i)$ are independently and identically distributed (i.i.d.) with the joint distribution $P_{LS}(x,y):=\bP{L \leq x, S \leq y}$ for each $i=1,\ldots, N$. The corresponding joint probability density function is given by $p_{LS}(x,y) = \frac{\partial^2}{\partial x \partial y} P_{LS}(x,y)$. We assume that the marginal densities $p_L(x)$ and $p_S(y)$ are continuous on their support. $L_{\textrm{min}}$ and $S_{\textrm{min}}$ denote the minimum values for load $L$ and free-space $S$, respectively, throughout the paper. We assume that $L_{\textrm{min}},  S_{\textrm{min}} > 0$.

Our load redistribution model is intimately related to the {\em democratic} fiber bundle model \cite{AndersenSornetteKwan,Daniels1945} where $N$ parallel fibers with failure thresholds $C_1, \ldots, C_N$
share an applied total force $F$ {\em equally}. In this line of literature, it has been of interest to study the dynamics of recursive failures in the bundle as the applied force
$F$ increases; e.g., see \cite{sornette1997conditions,pradhan2003failure,roy2015fiber}. This model was used by \cite{PahwaScoglioScala} in the context of power line networks, with $F$ corresponding to the total load shared {\em equally} by $N$ power lines. See also \cite{YaganNSR} for the latest developments on the democratic fiber bundle model relating to a power line network. The relevance of the equal load-redistribution model for power systems stems from its relation to Kirchhoff's law in the mean-field sense. Our current partial load redistribution model builds upon that in \cite{YaganNSR}. Even though our model involves a {\em global} redistribution of a fraction of the extra load due to line failures, the capability to partially capture the failures is, in effect, due to the ability to keep a portion of the extra load due to failed lines in a {\em local} level, possibly in a secondary network. While our current work does not consider the interaction of local and global behavior, our framework helps build a step towards this direction.  

 {\bf Problem definition.}
Our main goal is to study the robustness of the flow network under the partial load redistribution rule described above. We consider a {\em random} attack or a random failure that leads to $p$-fraction of the lines to be removed from the system. We assume that the load of these initially failed lines are redistributed {\em in full} to the non-attacked lines, with each non-attacked line receiving an equal portion of the total load failed. Our motivation in distinguishing these initial failures (resulting from an attack) from failures due to overloading of lines is two-fold. Firstly, in the case of a physical attack to the system, we would expect any advanced circuitry or inter-connection to other networks to be damaged along with the failed lines, making it impossible for $\varepsilon$-fraction of the failed load to be absorbed. Secondly, this assumption ensures that a random attack against $p$-fraction of the lines is equivalent (in the mean-field sense) to a disturbance caused by increasing the initial load of every line (or, force applied to every fiber) by $p\mathbb{E}[L]/(1-p)$. This in turn enables our analysis to provide insights on the robustness of the system against both  random attacks (as commonly considered in the context of power systems) as well as the increase of total applied load or force (as commonly considered in the context of fiber bundles).

After the initial load  redistribution, the amount of load on each alive line will be given by its initial load plus its share of the total load of the failed lines. This, in turn, leads to the failure of additional lines due to the updated flow exceeding their capacity. In the ensuing stages of this process, the network is assumed to have the capability of {\em absorbing} (i.e., removing from the system) $\varepsilon$-fraction of the load from lines who fail due to overloading. Put differently, if a line $\mathcal{L}_i$ fails due to its load $L_i(t)$ at time $t$ exceeding its capacity, then only $(1-\varepsilon)L_i(t)$ amount of load will be redistributed, in an equal manner, to the remaining lines; as mentioned before, the system is assumed to absorb the portion $\varepsilon L_i(t)$ either by help of advanced circuitry or by means of shedding that portion of load.

In the most general scenario for partial redistribution, we could allow $\varepsilon$ to depend on time $t$ and the extra load incurred per line at that time. However, our main goal for analysis in this paper is to understand the case when $\varepsilon$ is constant throughout time. The load redistribution process continues recursively until no further failures occur, potentially generating a {\em cascade of failures}. Our goal is to understand the limits associated with this process. We let $n_{\infty}(p,\varepsilon)$  denote the {\em final} (i.e., steady-state) fraction of alive lines when a $p$-fraction of lines is randomly attacked initially; as before $\varepsilon$ denotes the fraction of flow from {\em overloaded} lines that will be lost (and thus will not be redistributed to the remaining lines) at each stage of the cascade process. 

We derive expressions for $n_{\infty}(p,\varepsilon)$ for all attack sizes $0<p<1$ and any $0 \leq \varepsilon \leq 1$ in order to understand the {\em robustness} of the network under the partial load redistribution model. We will be particularly interested in understanding the {\em critical} attack size $p^{\star}$ at which $n_{\infty}(p,\varepsilon)$ drops to zero, and in developing design guidelines (in the sense of choosing the distribution $p_{LS}$) to optimize network robustness under given constraints.

With regard to the notation in use: 
Probabilistic statements are made with respect to  probability measure $\mathbb{P}$, and we
denote the corresponding expectation operator by $\mathbb{E}$. The indicator function of an event $A$ is denoted by $\1{A}$. 

\section{Analytic Results}
\label{sec:Results}

In this section, we provide a mean-field analysis for the cascading failures of lines under the model described in Section \ref{sec:Model}. We start by deriving recursive relations concerning the fraction $f_t$ of lines that are {\em failed} at time stage $t=0,1,\ldots$. The number of links that are still alive at time $t$ is then given by $N_t = N (1-f_t)$ for all $t=0,1,\ldots$. The cascading failures start with a random attack that targets a fraction $p$ of  lines. Hence, we have $f_0 = p$. Upon the failure of these $f_0$ lines, their load will be redistributed
to the remaining $(1-f_0)N$ lines, with each remaining line receiving an equal portion of the failed load. Since the $pN$ lines that have been attacked are selected uniformly at random, the mean total load that will be redistributed to the remaining lines is given by $\bE{L} p N$. The resulting extra load per alive line, $Q_0$, is thus given by 
\begin{equation}
Q_0 =  \frac{\bE{L} p N }{(1-p) N} = \bE{L} \frac{f_0}{1-f_0}.
\label{eq:Q_0}
\end{equation}

In the next stage, a line $i$ that survives the initial attack fails when its new load reaches its capacity. For convenience, we assume that a line also fails when its load equals its capacity, i.e., when
\[
 L_i + Q_0 \geq L_i + S_i, 
\]
or, equivalently $S_i \leq Q_0$. Therefore, at stage $t=1$, an additional fraction $\bP{S \leq Q_0}$ of the lines that were alive at the end of stage 0 fail. This yields
\[
f_1= f_0 + (1-f_0) \bP{S \leq Q_0} = 1-(1-f_0) \bP{S > Q_0}.
\]

Now, to compute the extra load per alive line at stage $1$, which we denote by $Q_1$, we need to sum the total load of the lines failed precisely at  stage $1$, multiply it by $(1-\varepsilon)$ to account for the load absorbed or lost as explained in Section \ref{sec:Model}, divide it by the new system size $(1-f_1)N$ and add this fraction on top of the existing extra load. With $\mathcal{A}$ denoting the initial set of lines attacked, this gives
\begin{align*}
 Q_1 
&= Q_0 +\frac{1-\varepsilon}{(1-f_1)N} \cdot \bE{\sum_{i \not \in \mathcal{A}: S_i \leq Q_0} (L_i+Q_0)} \\
&= Q_0   + \frac{1-\varepsilon}{(1-f_1)N} \cdot  \sum_{i\not \in \mathcal{A}} \bE{(L_i+Q_0) \1{S_i \leq Q_0}} \\
&= Q_0  + (1-\varepsilon)(1-f_0) \frac{\bE{(L+Q_0) \cdot \1{S \leq Q_0}}}{1-f_1} ,
\end{align*}
where the last step uses $|\mathcal{A}|/N=p=f_0$. 

At the stage $t=2$, the following two conditions are needed for a line to still stay alive: i) it should not have failed until this stage, which happens with probability $1-f_1$ and necessitates its free-space to satisfy $S > Q_0$; and ii) its free-space should also satisfy $S > Q_1$ so that its capacity is still larger than its current load. Thus, the fraction of failures $f_2$ at this stage is given by
\[
f_2 = 1 - (1-f_1) \bP{S > Q_1 ~|~ S > Q_0}.
\]
On the other hand, we can calculate the total load that is redistributed to the remaining lines as before:
\begin{align*}
Q_2 &= Q_1 + (1-\varepsilon)(1-f_0)\frac{ \bE{(L + Q_1) \cdot \1{ Q_0 < S \leq Q_1}}}{1-f_2}.
\end{align*}

The form of the recursive equations for each $t=0,1, \ldots$ can now be seen to be as follows:
\begin{equation}
\begin{array}{ll}
 f_{t+1} = &1- (1-f_t) \bP{S > Q_t ~\bigg|~ S > Q_{t-1}}\\ 
 Q_{t+1} =&Q_t + (1-\varepsilon)(1-f_0)\frac{\bE{(L+Q_t) \cdot \1{Q_{t-1} < S \leq Q_t}}}{1-f_{t+1}}\\
 N_{t+1} =&(1-f_{t+1}) N      
\end{array}
\label{eq:recursion}
\end{equation}
where $f_0 = p$, $N_0 = N (1-p)$, and $Q_0= \bE{L} \frac{p}{1-p}$. For convenience, we also
let $Q_{-1} = 0$. From (\ref{eq:recursion}) we see that cascades stop and a steady state is reached, i.e., $N_{t+2}=N_{t+1}$, if 
\begin{equation}
\bP{S > Q_{t+1} ~\bigg|~ S > Q_{t}} =1.
\label{eq:cond_steady_state}
\end{equation} 

We now work towards simplifying the recursion on $Q_t$ in order to obtain a better understanding of the condition 
(\ref{eq:cond_steady_state}) needed for cascading failures to stop. To this end,
we apply the first relation in (\ref{eq:recursion}) repeatedly to see that
\begin{eqnarray}\nonumber
\begin{array}{ll}
1-f_{t+1} &= (1-f_t) \bP{S > Q_t ~|~ S > Q_{t-1}} \\
1-f_{t} &= (1-f_{t-1}) \bP{S > Q_{t-1} ~|~ S > {Q_{t-2}}} \\
~~~\vdots & \\
1-f_{2} &= (1-f_{1}) \bP{S > Q_{1} ~|~ S > {Q_{0}}} \\
1-f_{1} &= (1-f_0) \bP{S > Q_0}. 
\end{array}
\end{eqnarray}
Applying these recursively starting from the last equality, we find that
\[
1-f_{t+1}  = (1-f_0) \prod_{\ell = 0}^{t} \bP{S > {Q_{\ell}} ~|~ S > {Q_{\ell-1}}},
\]
where we set $Q_{-1}=0$ as before. Since $Q_t$ is monotone increasing in $t$, i.e., $Q_{t+1}\geq Q_t$ for all $t$, we further obtain
\begin{align}
\lefteqn{1-f_{t+1}} &
\nonumber \\
 & =  (1-f_1) \frac{\bP{S > Q_{t}}}{\bP{S > Q_{t-1}}} \cdot \frac{\bP{S > Q_{t-1}}}{\bP{S > Q_{t-2}}} \cdots \frac{\bP{S > Q_{1}}}{\bP{S > Q_{0}}} 
   \nonumber \\
  & = (1-p)\bP{S > Q_{t}} 
  \label{eq:simplified_f_t}
\end{align}
This last expression confirms the intuitive result that the fraction of alive lines at stage $t+1$ is simply given by the fraction of lines who survive the initial attack and have more free-space than the extra load $Q_t$ that is distributed on every alive line at stage $t$. 

Using (\ref{eq:simplified_f_t}) in (\ref{eq:recursion}), it is now understood that the dynamics of cascading failures is fully governed and understood by the recursions on $Q_t$ given by 
\begin{align}
Q_{t+1} =Q_t + (1-\varepsilon)\frac{\bE{(L+Q_t) \cdot \1{Q_{t-1} < S \leq Q_t}}}{\bP{S > Q_{t}}}
\label{eq:new_osy_Q_t}
\end{align}
for each $t=0,1, \ldots$ (with $Q_0$ given at (\ref{eq:Q_0})), with the condition for reaching the steady-state still being  (\ref{eq:cond_steady_state}). Let $t^{\star}$ be the stage at which steady-state is reached, i.e., the first $t$ for which (\ref{eq:cond_steady_state}) holds. Then, the final system sizes $n_{\infty}(p,\varepsilon)$ defined as the fraction of alive lines at the steady state can be computed simply from (viz.~(\ref{eq:simplified_f_t}))
\begin{align}
    n_{\infty}(p,\varepsilon) = (1-p) \bP{S >Q_{t^{\star}}}.
    \label{eq:final_size_osy}
\end{align}
Throughout, we will be particularly interested in the critical attack size $p^{\star}$ defined as the largest attack  that the system can sustain (in the sense of having a positive final size $n_{\infty}(p,\varepsilon)$); i.e., for given $\varepsilon$ we let
\[
p^{\star}(\varepsilon) = \sup\{p > 0:n_{\infty}(p,\varepsilon) > 0\}
\]

In order to demonstrate the impact of the absorption factor $\varepsilon$ together with the number of stages needed to reach a steady-state, we find it useful to further simplify (\ref{eq:new_osy_Q_t}). By simple algebra, we get
\begin{align}\nonumber
    & Q_{t+1}\bP{S > Q_{t}} 
    \\ \nonumber
    &= Q_t\bP{S > Q_{t}}  +(1-\varepsilon)Q_t\bP{Q_{t-1}<S\leq Q_t} 
    \\ \nonumber
    &~~~~~ +(1-\varepsilon)\bE{L\cdot \1{Q_{t-1} < S \leq Q_t}} 
    \\ \nonumber 
    &= Q_t\bP{S > Q_{t-1}} - \varepsilon Q_t \bP{Q_{t-1} < S \leq Q_t} \\ \nonumber
    &~~~~~ + (1-\varepsilon)\bE{L\cdot \1{Q_{t-1} < S \leq Q_t}}, 
\end{align}
which is equivalent to the following difference relation 
on the sequence $Q_{t+1}\bP{S > Q_{t}}$
\begin{align}\label{eq:rec2}
    &Q_{t+1}\bP{S > Q_{t}} - Q_{t}\bP{S > Q_{t-1}} 
    \\ 
    & = (1-\varepsilon)\bE{L \1{Q_{t-1} < S \leq Q_t}} - \varepsilon Q_t  \bP{Q_{t-1} < S \leq Q_t}
    \nonumber
\end{align}
We see from (\ref{eq:rec2}) that higher values of $\varepsilon$  have suppressing effect on the growth of $Q_t$, leading to steady-state being reached {\em faster} (i.e., in small number of steps), and with a larger final system size in view of (\ref{eq:final_size_osy}).

It is desirable to obtain a closed-form solution 
for $Q_{t^{\star}}$ by solving the difference equation (\ref{eq:final_size_osy}); in view of (\ref{eq:final_size_osy}) this would lead to a closed-form expression for the final system size
$n_{\infty}(p,\varepsilon)$.
However, applying
(\ref{eq:rec2}) recursively leads to a telescoping sum given by
\begin{align}\nonumber
    &\hspace{-0.09in}Q_{t+1}\bP{S > Q_{t}} - Q_0
    \\ \nonumber
    &=(1-\varepsilon)\sum_{i=0}^{t}\bE{L\cdot \1{Q_{i-1} < S \leq Q_i}} \\ &~~~~ - \varepsilon\sum_{i=0}^{t}Q_i\bP{Q_{i-1} < S \leq Q_i} 
\end{align}
It is now clear that unless $\varepsilon =0$ or $\varepsilon=1$, a direct expression for $Q_t$  (for arbitrary $t$) can not be obtained without going through the recursion (\ref{eq:rec2}) and obtaining each one of $Q_1, Q_2, \ldots, Q_{t-1}$. Therefore, it is also not possible to derive a closed-form expression for $Q_{t^{\star}}$ and $n_{\infty}(p,\varepsilon)$.



\section{Numerical results}
\label{sec:Simu}

In this section, we confirm our theoretical findings via   numerical simulations. We focus on three commonly known distributions for the load and free-space variables: i) Uniform, ii) Pareto, and iii) Weibull. The probability density functions corresponding to these distributions are given below for a generic random variable $L$.
\begin{enumerate}
\item Pareto Distribution: $L \sim {Pareto}(L_{\textrm{min}}, b)$. With $L_{\textrm{min}}>0$ and $b>0$, the support set is $x \geq L_{min}$ and the density is given by
\[
p_L(x) =  L_{\textrm{min}}^{b} b x^{-b-1}.
\]
We also enforce $b>1$ in order to ensure that $\bE{L}=b L_{\textrm{min}}/(b-1)$ is finite. The Pareto family distributions are also known as {\em power-law} distributions and have been extensively used in many fields.
\item Uniform Distribution: $L \sim U(L_{\textrm{min}},L_{\textrm{max}})$. The support set is $L_{min} \leq x \leq L_{max}$ and the density is given by
\[
p_L(x)=\frac{1}{ L_{\textrm{max}}  - L_{\textrm{min}} }
\]
\item Weibull Distribution: $L \sim Weibull(L_{\textrm{min}},\lambda, k)$. With $\lambda, k, L_{\textrm{min}}>0$, the support set is $x \geq L_{min}$ and the density is given by
\[
p_L(x) = \frac{k}{\lambda} \left(\frac{x-L_{\textrm{min}}}{\lambda} \right)^{k-1} e^{-\left(\frac{x-L_{\textrm{min}}}{\lambda} \right)^{k}}.
 \]
The case $k=1$ corresponds to the exponential distribution, and $k=2$ corresponds to Rayleigh distribution. The mean load is given by $\bE{L}=L_{\textrm{min}}+ \lambda \Gamma (1+1/k)$, where $\Gamma(\cdot)$ is the gamma-function $\Gamma(x)=\int_{0}^{\infty} t^{x-1} e^{-t} dt$.
\end{enumerate}

\begin{figure}[!t]
\centering{
\hspace{-0.5cm} 
\includegraphics[totalheight=0.3\textheight]{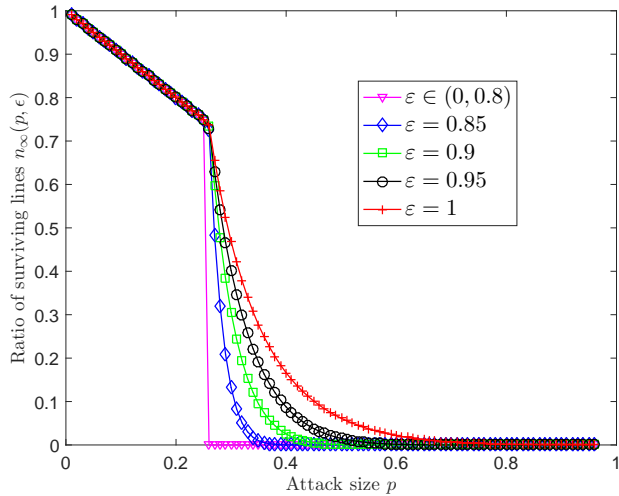}}
\caption{\sl The breakdown of the system is demonstrated, where $L_1, \ldots, L_N$ are drawn 
from a Pareto distribution with $b=2$, $L_{\textrm{min}}=10$. We also have $S=0.7L$ so that $S_1, \ldots, S_N$ are such that $S_i=0.7L_i$. We plot the ratio of surviving lines $n_{\infty}(p,\varepsilon)$ as a function of the attack size $p$ for various $\varepsilon$ values.}
\label{fig:1_step} 
\end{figure}

\begin{figure}[!t]
\centering{
\hspace{-0.5cm} 
\includegraphics[totalheight=0.3\textheight]{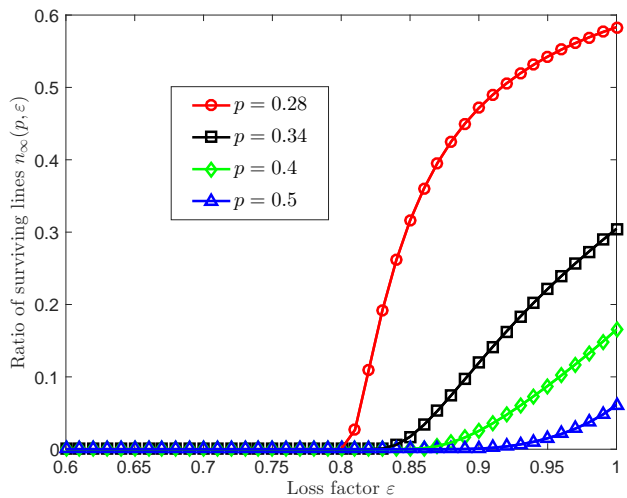}}
\caption{\sl In the setting of Fig. \ref{fig:1_step}, we plot the relative final size $n_{\infty}(p,\varepsilon)$ as a function of the loss/absorption factor parameter $\varepsilon$ for different attack sizes $p$.}
\label{fig:step2} 
\end{figure}

Next, we confirm our results presented in Section \ref{sec:Results} concerning the response of the system to attacks of varying sizes; i.e., concerning the final system size $n_{\infty}(p,\varepsilon)$. We are particularly interested in the transition behavior around the critical attack size $p^{\star}$. In all simulations, we fix the number of lines at $N=10^6$, and for each set of parameters being considered (e.g., the distribution $p_{LS}(x,y)$ and attack size $p$) we run $200$ independent experiments. In all figures below, the symbols represent the {\em empirical} value of the final system size $n_{\infty}(p,\varepsilon)$ (obtained from simulations by averaging over $200$ independent runs for each data point), and solid lines represent the analytic results computed from (\ref{eq:final_size_osy}) with $Q_{t^{\star}}$ obtained by iterating (\ref{eq:new_osy_Q_t}) while checking the condition (\ref{eq:cond_steady_state}) at each iteration step.

We start our numerical results with Pareto distribution. In Figs. \ref{fig:1_step} - \ref{fig:step2}, we let $L_1, \ldots, L_N$ be drawn from Pareto distribution with $b=2$, $L_{\textrm{min}}=10$ and $S=0.7L$. In Fig. \ref{fig:1_step}, we plot the ratio of surviving lines $n_{\infty}(p,\varepsilon)$ as a function of the attack size $p$ for various $\varepsilon$ values. We already know from the analysis in \cite{YaganPRE} that for $\varepsilon=0$, Pareto distribution always fosters an abrupt first-order transition behavior at $p^{\star}$. We observe that this first order transition behavior continues to hold as $\varepsilon$ is increased from $0$ to $0.8$. In fact, up until that point, the system's ability to absorb $\varepsilon$-fraction of the failed load  at each stage does not affect the final system size $n_{\infty}(p,\varepsilon)$. 
Only after $\varepsilon=0.8$ the behavior of $n_{\infty}(p,\varepsilon)$ starts to change as shown in Fig. \ref{fig:1_step}. For a complementary visualization, we plot the behavior of the ratio of surviving node $n_{\infty}(p,\varepsilon)$ as a function of $\varepsilon$ for different values of initial attack size $p$. We see that the transition no longer fosters any sharp behavior and continuously improves as it reaches $\varepsilon=1$. We next provide the mathematical justification for this observation.

When $\varepsilon=1$, an overloaded line will be removed from the system {\em without} its load being redistributed to the remaining lines. Put differently, the only time redistribution will take place is stage 1, where $Q_0$ gets redistributed to each of the $(1-p)N$ lines that have not been attacked. Therefore, a line that is not in the initial attack will be included in the final system size as long as its free-space is larger than $Q_0$. This leads to having 
\begin{align}
    n_{\infty}(p,\varepsilon) = (1-p)\bP{S>Q_0}
\end{align}
where $Q_0 = \frac{p}{1-p}\bE{L}$. Therefore, $n_{\infty}(p,\varepsilon)$ is continuous in $p$ whenever the marginal distribution of $S$ is continuous, a condition satisfied in all distributions considered in the numerical results. Therefore, in the case of perfect $\varepsilon$, $n(p,\varepsilon)$ hits zero and the system fully collapses in a continuous fashion. We further note that the distribution of $S$ under fixed $\bE{L}$ and $\bE{S}$ can be selected so that the critical attack size $p^{\star}$ can be made arbitrarily close to $1$. Selecting $S$ as a binary random variable at $\{s_1,s_2\}$ with $s_1<\bE{S}<s_2$ independent from $L$ such that $p_1s_1+p_2s_2=\bE{S}$ is sufficient to see this phenomenon. For fixed $s_1<\bE{S}$, one can select $s_2$ arbitrarily large with $p_1,p_2 \neq 0$ such that $p_1s_1+p_2s_2=\bE{S}$ and this proves that $\bP{S>Q_0}$ can be made non-zero {\em irrespective of the value of the initial attack size $p$}, i.e., $p^{\star}$ can be made arbitrarily close to 1.


\begin{figure}[!t]
\centering{
\hspace{-0.5cm} 
\includegraphics[totalheight=0.3\textheight]{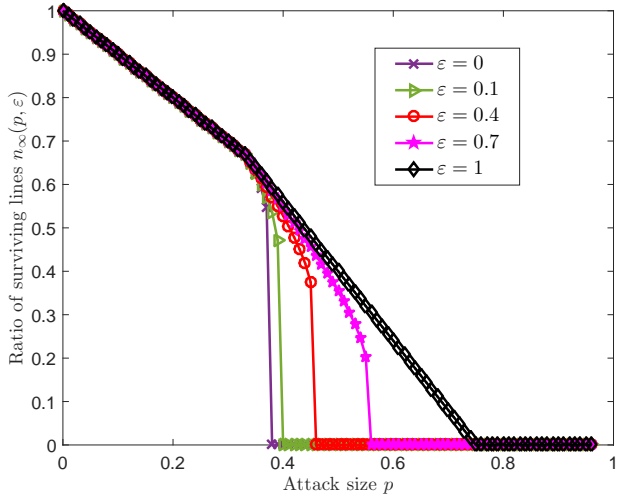}}
\caption{\sl The breakdown of the system is demonstrated, where $L_1, \ldots, L_N$ are drawn independently from a uniform distribution with $L_{\textrm{min}}=10$ and $\mathbb{E}[L] = 20$, and 
$S_1, \ldots, S_N$ are drawn independently from a  uniform distribution with $S_{\textrm{min}} =10$ and  $\bE{S}=35$. We plot the relative final size $n_{\infty}(p,\varepsilon)$ as a function of the attack size $p$ for different $\varepsilon$.}
\label{fig:step1} 
\end{figure}

In Fig. \ref{fig:step1}, we plot the ratio of surviving lines $n_{\infty}(p,\varepsilon)$ with respect to $p$ when $L_1, \ldots, L_N$ are drawn from the load L and extra space S is independent and uniformly distributed with $L_{min} =S_{min} =10$ and $\mathbb{E}[L] = 20$ and $\bE{S}=35$. We observe that the transition behavior has a failure with preceding divergence and the critical threshold $p^{\star}$ migrates from $p^{\star}(\varepsilon)=0.375$ at $\varepsilon=0$ to $p^{\star}(\varepsilon)=0.75$.

\begin{figure}[!t]
\centering{
\hspace{-0.5cm} 
\includegraphics[totalheight=0.3\textheight]{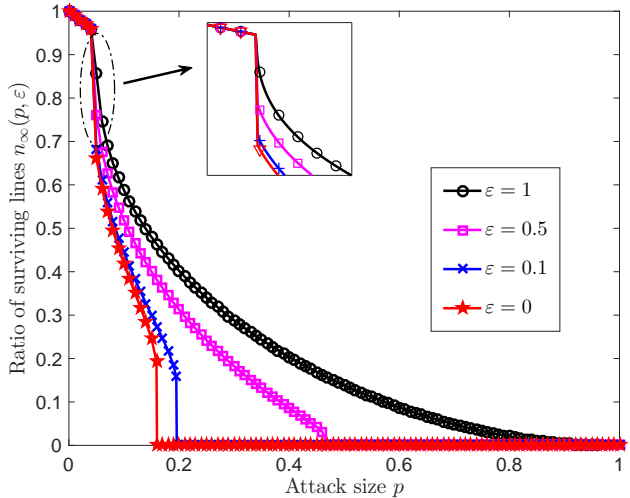}}
\caption{\sl The breakdown of the system is demonstrated, where $L_1, \ldots, L_N$ are drawn from a Weibull distribution with $k=0.4$, $\lambda=100$, $L_{\textrm{min}}=10$, and free-space for each lines is given by $S=1.74L$. We plot the ratio of surviving lines $n_{\infty}(p,\varepsilon)$ as a function of the attack size $p$.}
\label{fig:2_step} 
\end{figure}

In Fig. \ref{fig:2_step}, we plot the ratio of surviving lines $n_{\infty}(p,\varepsilon)$ with respect to $p$ when $L_1, \ldots, L_N$ are drawn from Weibull distribution with $k=0.4$, $\lambda=100$, $L_{\textrm{min}}=10$, $S=1.74L$. We see that the Weibull distribution gives rise to a richer set of possibilities for the transition of $n_{\infty}(p,\varepsilon)$. Namely, we see that an abrupt rupture, a rupture with preceding divergence as well as a first-order transition followed by a second-order transition that is followed by an ultimate first-order breakdown are all possible in this case. As the parameter $\varepsilon$ is increased, the transition behavior gets smoother. 

We finally examine the phase diagrams corresponding to cases presented above and reveal the emergence of tricritical points. In Fig. \ref{fig:phase}, we plot the loss factor $\varepsilon$ and attack size $p$ pairs for which the phase transition occurs in either first or second order. We denote a first order transition by straight line and a second order transition by dashed line. In Fig. \ref{fig:phase}, we refer to the case of $L \sim \text{Pareto(2,10)}$ and $S=0.7L$ as Case 1, the case of independent $L$, $S$ with $L \sim \text{Uniform([10,30])}$ and $S \sim \text{Uniform([10,60])}$ as Case 2 and the case of $L \sim \text{Weibull(10,100,0.4)}$ and $S=1.74L$ as Case 3. We observe that Case 1 fosters a very sharp phase diagram in that a first-order phase transition occurs at the same attack size for all loss factors smaller than a certain $\varepsilon$ value, and a second-order transition is seen after that $\varepsilon$ is exceeded. In contrast, Case 3 has a smoother phase diagram as the phase transition switches from first order to second order after a certain tricritical point \cite{parshani2010interdependent}. We also observe that the transition in Case 2 is always of first order and the corresponding phase diagram is smooth.  

\begin{figure}[!t]
\centering{
\hspace{-0.5cm} 
\includegraphics[totalheight=0.28\textheight]{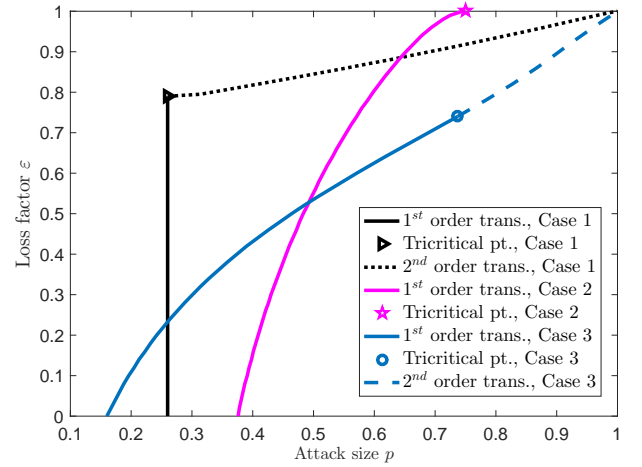}}
\caption{\sl The phase diagrams and the resulting tricritical behavior is shown for Cases 1, 2 and 3. For each value of the loss/absorption factor $\varepsilon$, we show the corresponding {\em critical} attack size $p$. Solid curves represent cases where the transition at the corresponding critical attack size is first-order, while dashed curves stand for cases where a second-order phase transition occurs.}
\label{fig:phase}   
\end{figure}

\section{Optimizing Robustness}
\label{sec:Optimal}

\subsection{Quantifying robustness}
A typical metric to assess network robustness is the percolation threshold $p^{\star}$ at which the system fully collapses as a result of the cascading failures \cite{YaganPRE,YaganNSR}. In particular, we know from earlier works \cite{YaganPRE,YaganNSR} that for $\varepsilon=0$, $p^{\star}_{optimal}=\frac{\mathbb{E}[S]}{\mathbb{E}[S] + \mathbb{E}[L]}$ and it is achieved by a Dirac delta distribution at $\bE{S}$. We observed in the numerical results in Section \ref{sec:Simu} that the presence of the parameter $\varepsilon$ changes $p^{\star}$. For example, we have seen that if more load is absorbed from the overloaded lines by increasing $\varepsilon$, $p^{\star}$ may increase or decrease with respect to the case $\varepsilon=0$. In cases when $p^{\star}$ increases $n_{\infty}(p,\varepsilon)$ decreases. It is, therefore, difficult to asses the robustness using only the $p^{\star}$ metric; e.g., see Fig.~\ref{fig:3_step} where the distribution that leads to the highest  $p^{\star}$ does not maximize the final system size $n_{\infty}(p,\varepsilon)$ uniformly across all attack sizes $0 \leq p \leq 1$.

In order to quantify the overall robustness of the network under all possible attack sizes, we consider a metric that measures the area under $n_{\infty}(p, \varepsilon)$ over $0 \leq p \leq 1$. Namely, we 
let
\begin{align}\label{area_metric}
    \mathcal{R}(\varepsilon) = \int_0^1 n_{\infty}(p,\varepsilon)\mathrm{d}p.
\end{align}
The metric $ \mathcal{R}(\varepsilon)$
was introduced in \cite{scheider2011},
and can be seen \footnote{The metric in \cite[Equation 1]{scheider2011} is introduced as the sum of surviving nodes over all possible numbers of attacked lines. One can show by the almost sure convergence of the probabilities to the fraction of line failures due to the law of large numbers and the bounded convergence theorem that the metric $\mathcal{R}(\varepsilon)$ in (\ref{area_metric}) represents the same metric in \cite{scheider2011} as the number of lines $N$ grows.}
 to represent the {\em expected} final system size in response to an attack whose size $p$ is selected uniformly at random over $[0,1]$. It is in this spirit that the metric $\mathcal{R}(\varepsilon)$ quantifies the overall system robustness under a range of attack sizes for fixed $\varepsilon$. Alternative metrics can also be defined where the attack size $p$ is drawn from an arbitrary distribution $F(p)$, e.g., to account for the fact that attacks of certain size might be more likely than others. In that case, we would again compute the mean of the final system size, i.e., 
\begin{align} \label{alternative}
  \mathbb{E}_p\left[n_{\infty}(p,\varepsilon)\right] = \int_0^1 n_{\infty}(p,\varepsilon)\mathrm{d}F(p)  
\end{align}
Clearly, $\mathcal{R}(\varepsilon)$ defined in (\ref{area_metric}) is recovered when $F(p)$ is the uniform distribution over $[0,1]$.

\subsection{Maximizing the robustness metric $\mathcal{R}(\varepsilon)$}

Our next result conclusively establishes that for any $\varepsilon$, the Dirac delta distribution \footnote{We note that Dirac delta distribution can be viewed as a limiting distribution for a family of continuous distributions and therefore it can be treated within the framework of our paper that assumed continuous $P_{LS}(x,y)$} 
for free-space $S$ optimizes the robustness of the system with respect to the metric $\mathcal{R}(\varepsilon)$ among all possible distributions $P_{LS}(x,y)$ with fixed $\bE{S}$ and $\bE{L}$. First, we note that if $p_{LS}(x,y) = p_L(x) \delta(y-\bE{S})$, then the final system size is independent of $\varepsilon$. This is because after the initial attack, either all lines will fail (if $\bE{S} \leq p\bE{L}/(1-p)$), or they will all survive and the cascades will not continue. Thus, the final system size under the Dirac-delta distribution of free-space is given \cite{YaganNSR} by
\begin{equation}
n_{\delta,\infty}(p) = \left \{  
\begin{array}{cc}
1-p  &  \textrm{if $p < p_{\delta}^{\star} $}   \\
0  &     \textrm{if $p \geq p_{\delta}^{\star} $}
\end{array}
\right.
\label{eq:robustness_of_dirac_S}
\end{equation}
where, the critical attack size $p_{\delta}^{\star}$ is given by
\[
p_{\delta}^{\star}=\frac{\mathbb{E}[S]}{\mathbb{E}[S] + \mathbb{E}[L]}.
\] 

We will show that the distribution $p_{LS}(x,y) = p_L(x) \delta(y-\bE{S})$ maximizes the robustness metric $\mathcal{R}(\varepsilon)$ among all $p_{LS}$ with mean values for $L$ and $S$ fixed at $\bE{L}$ and $\bE{S}$, respectively. 
In view of (\ref{eq:robustness_of_dirac_S}), this will follow if we show that
\begin{align}
  \int_0^1 n_{\infty}(p,\varepsilon, p_{LS})\mathrm{d}p \leq  \int_0^{p_{\delta}^{\star}} (1-p) \mathrm{d}p
  \label{eq:to_show1}
\end{align}
where $n_{\infty}(p,\varepsilon, p_{LS}(x,y))$ denotes the final system size under attack size $p$, when load and free-space values of the lines are generated independently from the distribution $p_{LS}(x,y)$ (with fixed $\bE{L}$, $\bE{S}$). 

From (\ref{eq:final_size_osy}) we note that  
\begin{align}
    n_{\infty}(p,\varepsilon, p_{LS}(x,y)) &\leq (1-p)\bP{S \geq Q_0}
    \nonumber
    \\ & = (1-p)\bP{S \geq \frac{p}{1-p}\bE{L}} ,
\end{align}
due to the fact that $Q_0 \leq Q_t$ for all $t=1,2, \ldots$. Therefore, we will get the desired result (\ref{eq:to_show1}) if we show that
\begin{align}\nonumber
    \int_0^1  (1-p)\bP{S \geq \frac{p}{1-p}\bE{L}} \mathrm{d}p \leq \int_0^{p_{\delta}^{\star}} (1-p) \mathrm{d}p,
\end{align}
or, equivalently that
\begin{align}\nonumber
    & \int_{p_{\delta}^{\star}}^1  (1-p)\bP{S \geq \frac{p}{1-p}\bE{L}} \mathrm{d}p 
    \nonumber \\
    & \leq \int_0^{p_{\delta}^{\star}} (1-p) \bP{S < \frac{p}{1-p}\bE{L}} \mathrm{d}p,
    \label{eq:to_show2}
\end{align}
Since $1-p$ is monotone decreasing over the range $0 \leq p \leq 1$, and both $\bP{S \geq \frac{p}{1-p}\bE{L}}$ and $\bP{S < \frac{p}{1-p}\bE{L}}$ are non-negative, (\ref{eq:to_show2}) will follow if we show that
\begin{align}\nonumber
     \int_{p_{\delta}^{\star}}^1 \bP{S \geq \frac{p}{1-p}\bE{L}} \mathrm{d}p 
    \leq \int_0^{p_{\delta}^{\star}}  \bP{S < \frac{p}{1-p}\bE{L}} \mathrm{d}p,
\end{align}
or, equivalently that
\begin{align}
\int_{0}^1 \bP{S \geq \frac{p}{1-p}\bE{L}} \mathrm{d}p \leq \int_0^{p_{\delta}^{\star}}\mathrm{d}p = 
\frac{\bE{S}}{\bE{S} + \bE{L}},
        \label{eq:to_show3}
\end{align}

In order to establish (\ref{eq:to_show3}), we  make a change of variables $x=\frac{p}{1-p}\bE{L}$ and write
\begin{align}\nonumber
    & \int_0^1 \bP{S \geq \frac{p}{1-p}\bE{L}} \mathrm{d}p \\ 
     \label{seq_2} 
    & = \int_0^{\infty} \bP{S \geq x} \mathrm{d}\left(\frac{x}{x+\bE{L}}\right) 
    \\ \nonumber & = \bP{S \geq x}\frac{x}{x+\bE{L}} \bigr|_{x=0}^{\infty} - \int_0^{\infty} \frac{x}{x+\bE{L}}\mathrm{d}(\bP{S \geq x}) \\  & = \bE{\frac{S}{S + \bE{L}}} \nonumber 
    \\ & \leq \frac{\bE{S}}{\bE{S} + \bE{L}}  \label{eq:bound}
\end{align}
where we use integration by parts in (\ref{seq_2}) and
 apply Jensen's inequality in (\ref{eq:bound}) for the  function $\frac{x}{x+\bE{L}}$  that is concave in $x$. This establishes (\ref{eq:to_show3}) and the desired result 
 (\ref{eq:to_show1}) follows in view of the preceding arguments.

This result shows that the system's robustness with respect to the metric $\mathcal{R}(\varepsilon)$ (defined at (\ref{area_metric})) is maximized under the constraints of fixed $\bE{L}$ and fixed $\bE{S}$ (and hence fixed $\bE{C}$), by giving each line an equal free-space $\bE{S}$, 
{\em irrespective of how the initial loads are distributed and the redistribution process in the later stages}. In other words, the robustness is maximized by choosing a line's capacity $C_i$ through $C_i = L_i + \bE{S}$ no matter what its load $L_i$ is. 

Note that in light of the results in Section \ref{sec:Simu}, the critical attack size $p^{\star}(\varepsilon)$ may be greater or less than $\frac{\bE{S}}{\bE{S} + \bE{L}}$ depending on the specific distribution $p_{LS}$ used (with fixed $\bE{L}$ and $\bE{S}$). In particular, we have seen that it is always possible to choose a marginal distribution with $\bE{S}$ such that $p^{\star}$ is arbitrarily close to $1$. Therefore, one might think that the area under $n_{\infty}(p,\varepsilon)$ while swiping all possible $p$ will be maximized when  $\varepsilon$ is increased and the system has the capability to absorb the extra load coming from the failing lines and eradicate the potentially detrimental effect of their failure to the overall system. Our result shows firmly that this intuition is incorrect and the metric $\mathcal{R}(\varepsilon)$ is instead maximized when the distribution of $S$ is the Dirac delta function centered at $\bE{S}$ (irrespective of
$\varepsilon$ and the  distribution of $L$).

We note that our argument follows from the facts that the extra load due to the initial attack $Q_0=\frac{p}{1-p}\bE{L}$ is monotone increasing, continuous, and convex in the initial attack size $p$. These properties are  expected to hold in a large set of instances of this problem. Therefore, the optimality of the Dirac-delta distribution of $S$ is likely to hold under more general cases where these properties hold. For instance, if the prior randomness $F(p)$ on the attack size $p$ in (\ref{alternative}) has a monotone decreasing derivative, i.e., if higher attack sizes are {\em less likely}, then the Dirac-delta distribution of free-space $S$ is still optimal with respect to the resulting metric in (\ref{alternative}). Such cases occur in situations where the malicious attacker is more likely to choose smaller attack sizes due to resource or time constraints. We similarly observe that if the support of the prior distribution of $p$ is contained in the interval $[0,\frac{\bE{S}}{\bE{S}+\bE{L}}]$, then the Dirac-delta function at $\bE{S}$ is optimal with respect to $\mathcal{R}(\varepsilon)$. Finally, the optimality of the Dirac-delta distribution of $S$ (in the sense of maximizing $\mathcal{R}(\varepsilon)$) holds irrespective of the value of $\varepsilon$. As such, this optimality prevails under any time variation in $\varepsilon$ or possible dependence of $\varepsilon$ on the instantaneous extra load per line $Q_t$.

While it is hard to generalize the optimality of Dirac delta function for $0<\varepsilon<1$ and any prior distribution on $p$, we observe that if $p$ is deterministic and $\varepsilon=1$, then the robustness metric in (\ref{alternative}) is given by
\begin{align}
   \mathbb{E}_p\left[n_{\infty}(p,\varepsilon)\right] =  n_{\infty}(p,1) = (1-p)\bP{S>Q_0}.
\end{align}
Then, in order to maximize $n_{\infty}(p,\varepsilon)$, it suffices to minimize $\bP{S>Q_0}=\bP{S>p\bE{L}/(1-p)}$ subject to $\bE{S}$ being fixed. If the known attack size satisfies $p\leq \frac{\bE{S}}{\bE{S}+\bE{L}}$, then this is achieved when the distribution of $S$ is given by the Dirac-delta function centered at $\bE{S}$. If, on the other hand, we have $p > \frac{\bE{S}}{\bE{S}+\bE{L}}$, then the optimal distribution of $S$ consists of two Dirac delta functions centered at $0$ and $\frac{p}{1-p}\bE{L}$, respectively, with appropriate probabilities selected such that the mean value is $\bE{S}$. In other words, if the attack size is larger than what can be resisted by giving each line an equal amount $\bE{S}$ of free-space, then it is optimal to give a fraction of lines {\em zero} free-space, while giving each of the other lines an equal amount of $\frac{p}{1-p}\bE{L}$ free-space; i.e., just enough so that they can handle the additional load of $Q_0$ that will be distributed to them after the attack.

\begin{figure}[t]
\centering{
\hspace{-0.55cm} 
\includegraphics[totalheight=0.29\textheight]{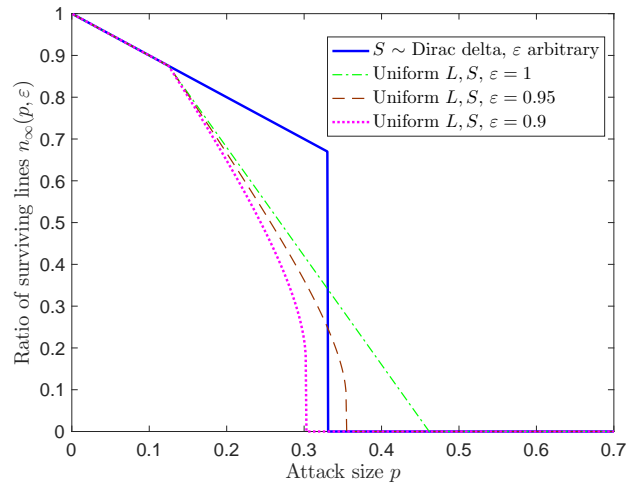}}
\caption{\sl We plot the ratio of surviving lines $n_{\infty}(p,\varepsilon)$ as a function of the attack size $p$ in the setting when $L$ and $S$ are independent and distributed uniformly as $U[0,120]$ and $U[10,60]$, respectively, as well as the setting when $S$ is Dirac delta distributed as $\delta(x-\bE{S})$.}
\label{fig:3_step} 
\end{figure}

\subsection{Simulations with synthetic and real-world data}
\label{subsec:real-data}

In Fig.~\ref{fig:3_step}, we plot the fraction of surviving lines $n_{\infty}(p,\varepsilon)$ as a function of the attack size $p$ for several $\varepsilon$ values, in the setting when $L$ and $S$ are independent and  uniformly distributed as $U[0,120]$ and $U[10,60]$, respectively. We compare the corresponding final system size $n_{\infty}(p,\varepsilon)$ when $S$ is Dirac delta distributed as $\delta(x-\bE{S})$. We observe that the area under $n_{\infty}(p,\varepsilon)$ is larger under the Dirac delta distribution compared to other cases. In Fig. \ref{fig:example}, we provide a comparison of the metric $\mathcal{R}(\varepsilon)$ for a family of Weibull distributions with fixed $\bE{L}$ and $\bE{S}$ while varying the scale parameter $k$ of the distribution. We observe that $\mathcal{R}(\varepsilon)$ is monotone increasing in $k$ and is maximized in the limiting case $k \rightarrow \infty$. This observation is in perfect agreement with our result given that the Weibull distribution approaches to a Dirac delta function as $k$ goes to infinity.

\begin{figure}[!t]
\centering{
\hspace{-0.85cm} 
\includegraphics[totalheight=0.31\textheight]{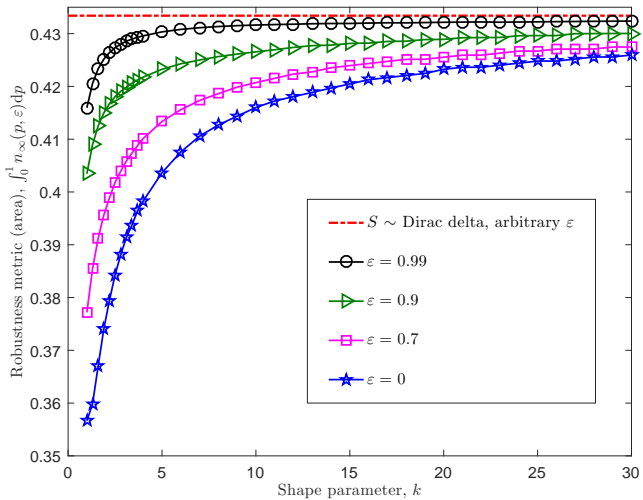}}
\caption{\sl $L_1, \ldots, L_N$ are drawn 
from Weibull distribution with $L_{\textrm{min}}=1$ and $k,\lambda$ such that $\bE{L}=2$ and $S=1.74L$. We plot the metric $\mathcal{R}(\varepsilon)$ in (\ref{area_metric}) as a function of the parameter $k$.}
\label{fig:example} 
\end{figure}

We now test our analytic results concerning the optimization of robustness on real-world power system data from the UK National Grid \cite{link}. In particular, we pick $N=75$ loads $L_1, \ldots, L_{75}$ from the Great Britain two degree power flow diagram for 2017/18 shown explicitly in Figure C.1 in \cite[ Appendix C]{link}. These power flows represent the power units in MVA that flow through lines in the utility grid in the scale shown in \cite[ Appendix C]{link}. 

Our goal is to compare the robustness of this power system, as quantified through the area metric, under different allocations of line capacities. In particular, we will compare the performance of the equal free-space allocation with respect to the schemes described below:
\begin{itemize}
    \item Proportional free-space allocation: For each line $i$, we set $S_i=\alpha L_i$, where $\alpha$ is the {\em tolerance} factor, set to be the same across all $N$ lines. 
    \item Equal capacity allocation: $S_i + L_i = C$ for all $i$ if $L_i < C$ and $S_i=0$ otherwise.
\end{itemize}
In order for this comparison to be fair, the total free-space available to all 75 lines will be fixed; since loads are already given from the UK Power-grid data set, this will amount to having the total capacity of 75 lines fixed. 
The total load of the 75 lines given in this data set is $60221$ power units. We assume that a total free-space of $168620$ power units is available to these lines and will be allocated according to one of three free-space allocation schemes to be considered. For the proportional free-space allocation, this leads to using a tolerance-factor of $\alpha=2.8$ for all lines, while for the equal capacity allocation scheme, this leads to fixing the capacity of all lines at $C = 3051$ power units. In the case of equal free-space allocation, this amounts to assigning the capacity of every line at their load plus $168620/75\simeq 2248$ units of free-space. These three schemes are compared with respect to the resulting area robustness metric defined here as
\begin{align} \label{eq:robmetric}
    \mathcal{R}(\varepsilon) = \frac{1}{N}\sum_{i=1}^{N}n_{\infty}(p_i,\varepsilon)
\end{align}
where $p_i=\frac{i}{N}$. 

The results are presented in Fig. \ref{fig:real}, where each data point corresponds to the average over $5000$ independent experiments. We observe that the equal free-space allocation performs significantly better than the other two capacity allocation schemes, in the sense leading to a larger value of the robustness metric (\ref{eq:robmetric}). For instance, when a random attack occurs whose scale $p$ is selected uniformly at random from $(0,1)$, we see that close to 50\% of the lines are expected to survive under the optimal choice of equal free-space allocation. We remind that under (\ref{eq:robmetric}), the metric $\mathcal{R}(\varepsilon)$ is  always less than $0.5$, meaning that the performance under the equal free-space allocation is close to the theoretically possible maximum robustness level.
This statement holds irrespective of the absorption factor $\varepsilon$ and in particular even if the system has no capability of absorbing the failed load. However, if we use the widely assumed setting of free-space being proportional to load, i.e., $C_i = (1+\alpha)L_i$, then we can expect only about 35\% of the nodes to survive the same scenario if the system has no load absorption capability. We see that even if the system is designed with full load absorption capability (i.e., $\varepsilon=1$), the expected number of surviving nodes increases only to about 41\%, still worse than the optimal case. The equal-capacity allocation leads to a higher robustness than the proportional free-space case, but the achieved robustness is still inferior to the case of where all lines have the same free-space. It is only when the system has almost perfect flow absorption capability (i.e., $\varepsilon =1$) that the performances of equal-capacity and equal free-space allocations become similar; for small $\varepsilon$ we see that equal free-space is significantly better. Although not reported here for brevity, our extensive simulation results indicate that similar conclusions apply for a wide range of $\alpha$ and $C$ values.

\begin{figure}[!t]
\centering{
\hspace{-0.95cm} 
\includegraphics[totalheight=0.26\textheight]{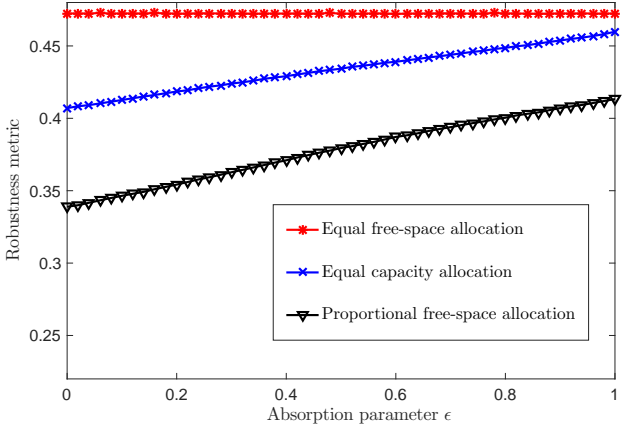}}
\caption{\sl Comparison of robustness performances of equal free-space allocation, equal capacity allocation and proportional free-space allocation for $N=75$ lines with load data obtained from the UK National Grid \cite[ Appendix C]{link}. We observe that the robustness is better for equal free-space allocation with respect to the other two benchmarks.}
\label{fig:real} 
\end{figure}

\subsection{Simulations under a  topology-based redistribution model}

The results presented in earlier sections of this paper were based on a model with a mean field assumption in that when a line fails, its load gets redistributed globally and equally among all active lines in the network. This mean field assumption may not hold for networks with possible local redistribution behavior such as power networks. On the other hand, it is well known that redistribution models with only local components fail to capture the long-range nature of Kirchhoff’s Law. Consequently, a model where the failed load is redistributed both locally and globally would be a more suitable one. To this end, we present in this section simulation results under a topology based redistribution model. We do so with an eye towards revealing whether the optimality of equal free-space distribution prevails when failed load is redistributed (at least in part) locally according to a network topology.  

\begin{figure}[!ht] 
    \centering
\subfigure[]
    {
    \includegraphics[totalheight=0.27\textheight]{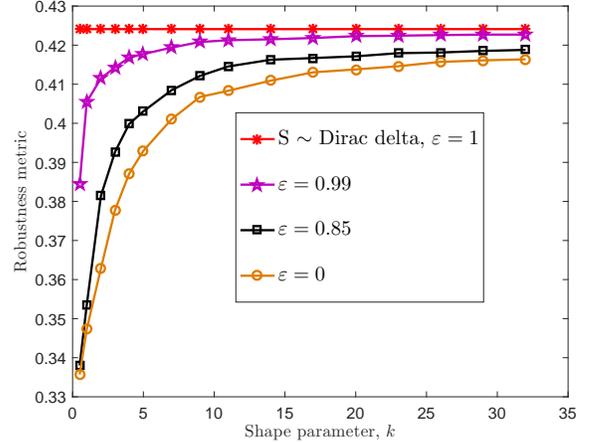}
    \label{fig:local1}} 
\subfigure[]
    {
    \includegraphics[totalheight=0.27\textheight]{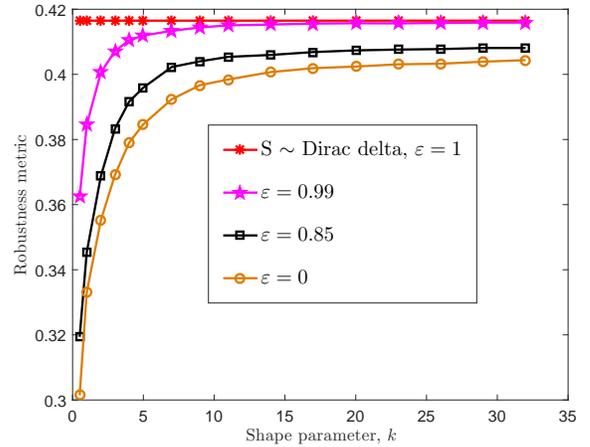}
    \label{fig:local2}} 
    \subfigure[]
    {
    \includegraphics[totalheight=0.27\textheight]{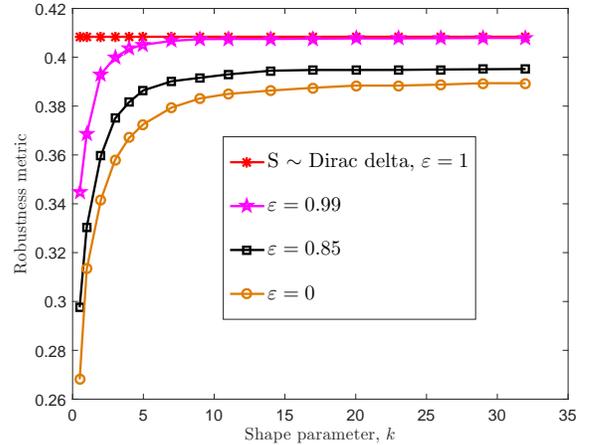}
    \label{fig:local3}} 
 \vspace{-4mm} 
 \caption{\sl We let $L_1, \ldots, L_N$ be obtained from Weibull distribution with $L_{\textrm{min}}=1$ and $k,\lambda$ such that $\bE{L}=2$ and $S=1.74L$. For 
 (a) $\gamma=0.25$, (b) $\gamma=0.6$, and (c) $\gamma=1.0$,
 the plots show the variation of the metric $\mathcal{R}(\varepsilon)$ in (\ref{eq:robmetric}) with the parameter $k$ for various $\varepsilon$ values. \vspace{-3mm}}
  \label{fig:local_together} 
\end{figure}

In the spirit of the redistribution models presented in \cite{MotterLai, Mirzasoleiman}, we consider a topology based redistribution model that combines local and global redistribution behaviors. This extended model gauges the locality of the redistribution by the parameter $\gamma \in [0,1]$ and the network topology is generated as an Erd\H{o}s-R\'enyi graph $\mathbb{G}(n,N)$. At each stage, the portion of the load that is not absorbed from each failed line is divided into two parts: $\gamma$-fraction is redistributed locally among neighboring lines (with each neighbor receiving an equal portion), and $(1-\gamma)$-fraction is redistributed equally among {\em all}  surviving lines (irrespective of topology). In this model, setting $\gamma=0$ recovers the mean-field model introduced in Section \ref{sec:Model}, while setting $\gamma=1$ gives a merely topology based redistribution model. 

For this new redistribution model, we have run simulations where we set the number of nodes as $n=250$ and number of edges as $N=8400$, and created a random network according to the Erd\H{o}s-R\'enyi $\mathbb{G}(n,N)$ model. This leads to each line having on average $132$ neighbors. The loads $\{L_1,\ldots,L_N\}$ are drawn from i.i.d. Weibull distribution (with $L_{\textrm{min}}=1$ and $k,\lambda$ such that $\bE{L}=2$) and the free spaces are set as $S_i=1.74L_i$. We run $100$ independent experiments for each parameter set, and report the average value of the robustness metric defined in (\ref{eq:robmetric}). 

\begin{figure}[!ht] 
    \centering
\subfigure[]
    {
    \includegraphics[totalheight=0.27\textheight]{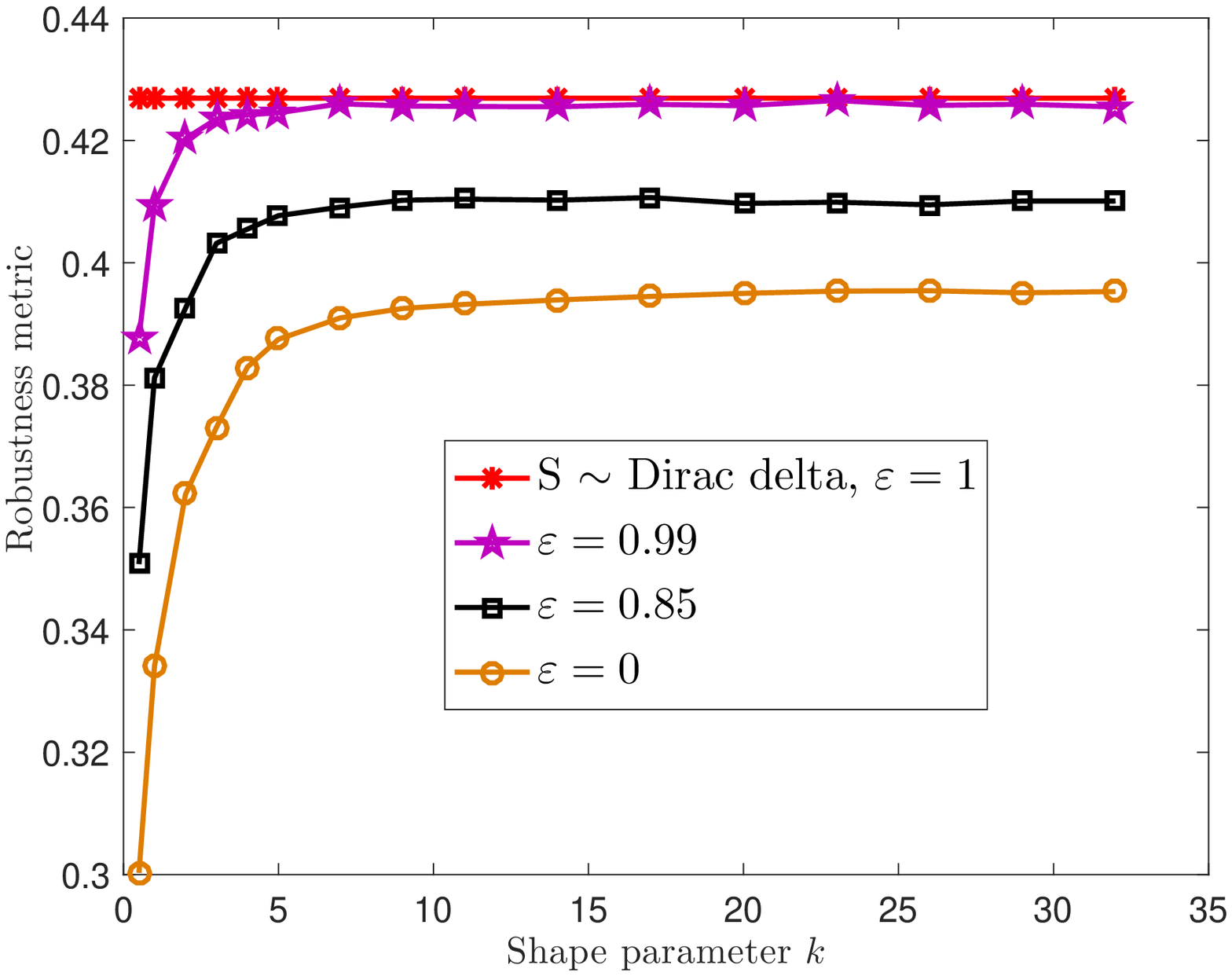}
    \label{fig:local21}} 
\subfigure[]
    {
    \includegraphics[totalheight=0.27\textheight]{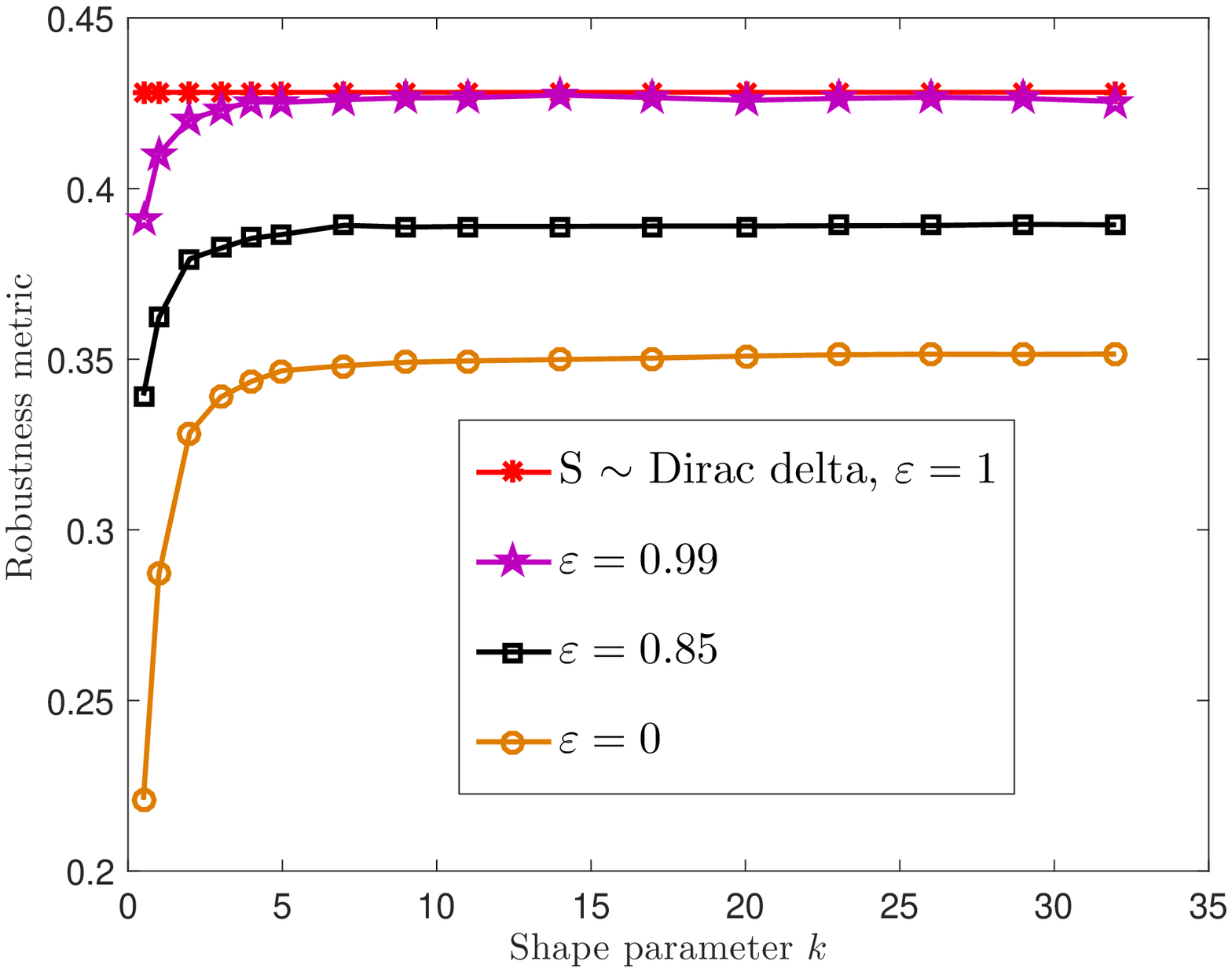}
    \label{fig:local22}} 
    \subfigure[]
    {
    \includegraphics[totalheight=0.27\textheight]{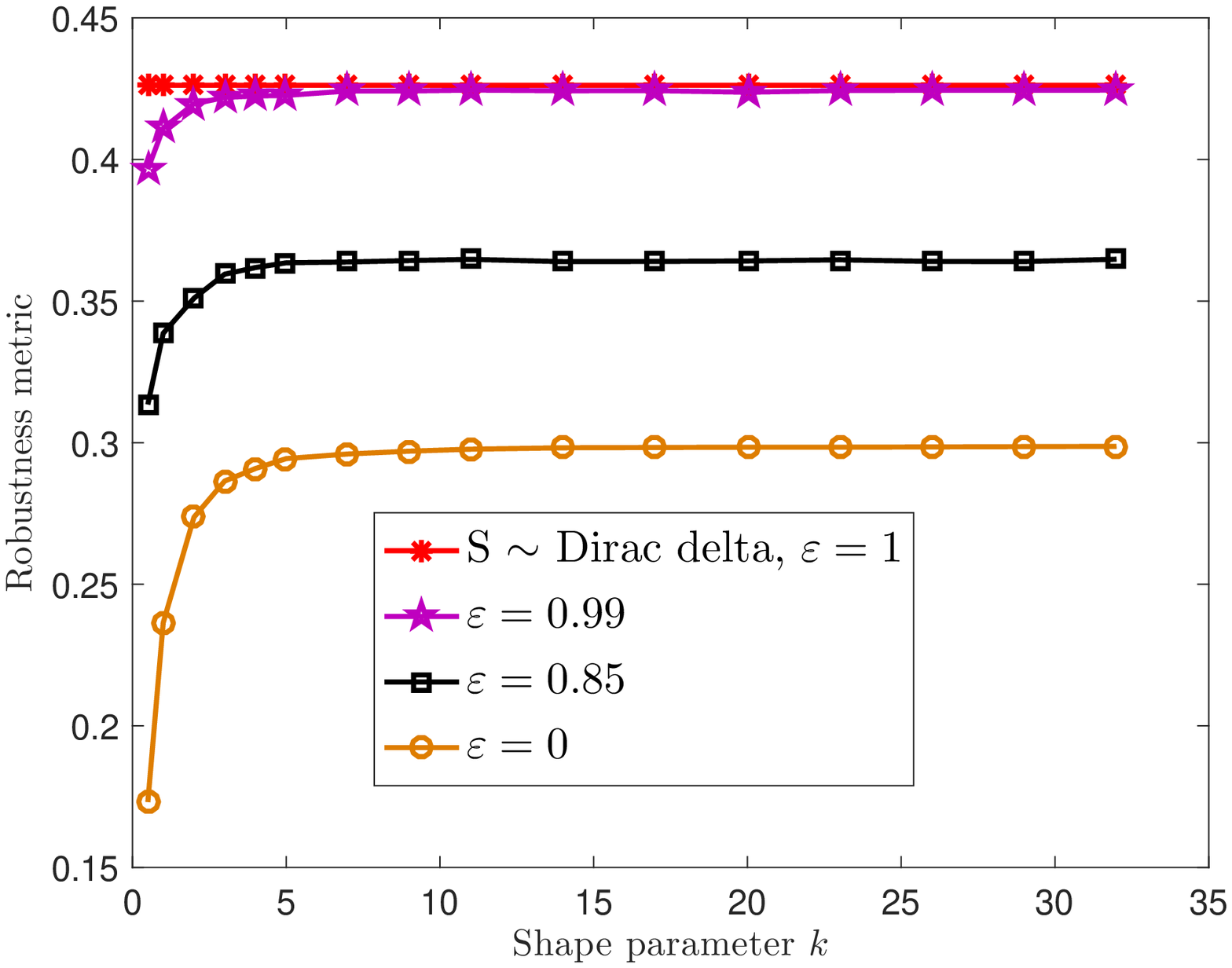}
    \label{fig:local23}} 
 \vspace{-4mm} 
 \caption{\sl We repeat the experiments in Fig. \ref{fig:local_together} for 
 (a) $\gamma=0.25$, (b) $\gamma=0.6$, and (c) $\gamma=1.0$,
 under a different Erd\H{o}s-R\'enyi graph with significantly smaller average number of neighbors. \vspace{-3mm}}
  \label{fig:local_together2} 
\end{figure}

The results are presented in Fig. \ref{fig:local1}, Fig. \ref{fig:local2} and Fig. \ref{fig:local3} for $\gamma=0.25$, $\gamma=0.6$ and $\gamma=1$, respectively; we remark that the setting in Fig. \ref{fig:example} corresponds to the case with $\gamma=0$. 
The most important observation is that Figs. \ref{fig:local1}-\ref{fig:local3} obtained under (partially) {\em local} load redistribution are qualitatively the same with the results given in Fig. \ref{fig:example} obtained under  mean-field analysis (i.e., global redistribution). For instance, we observe in all figures that as the parameter $k$ is increased, the robustness of the network increases monotonically with Dirac-delta distribution of free-space $S$ leading the maximum robustness. Thus, we conclude that the optimality of equal free-space allocation prevails even when the redistribution of the loads on failing lines are performed based on the topology of the network. Another observation we make from Figs. \ref{fig:local1}, \ref{fig:local2} and \ref{fig:local3} with comparison to Fig. \ref{fig:example} is that increasing the locality parameter $\gamma$ decreases robustness. This is in line with our understanding that introducing heterogeneity in the redistribution model causes further loss in robustness.

We finally test the effect of node degree on the robustness of the network. We consider the Erd\H{o}s-R\'enyi $\mathbb{G}(n,N)$ model with the same number of nodes $n=250$ and a significantly smaller number of edges $N=625$. In this case, each line has $8$ neighbors on average. The results are presented in Fig. \ref{fig:local21}, Fig. \ref{fig:local22} and Fig. \ref{fig:local23} for $\gamma=0.25$, $\gamma=0.6$ and $\gamma=1$, respectively. We observe the same qualitative behaviour in Fig. \ref{fig:local_together} in that as the shape parameter of the Weibull distribution gets larger, the robustness of the network improves irrespective of locality parameter $\gamma$ and the absorption parameter $\epsilon$. Additionally, we observe that robustness is affected by the connectivity of the network in different ways for different $\gamma$ and $\epsilon$. In particular, robustness significantly drops as the average number of neighbors decreases when $\epsilon$ is smaller while the effect of the number of neighbors on robustness is limited for higher $\epsilon$. This numerical observation points to an intricate relation among load absorption, the average number of neighbors and locality in the load redistribution process for their effects on survival and robustness against cascading failures. Overall, the numerical results in Figs. \ref{fig:local_together} and \ref{fig:local_together2} confirm the promise of uniformly allocating the free spaces to each node and gives reason to extend our results to more general network and redistribution models \footnote{Our intuition is that equal free-space allocation will be optimal under any model where the (load, free-space values) of lines are drawn independently from a common distribution. After all, it should be the higher connectivity or higher load of neighboring lines that might warrant allocating a line a higher free-space, instead of the load of the line itself as suggested by the linear free-space allocation scheme $S=\alpha*L$. This phenomenon already found presence and application in the context of information diffusion process, e.g., in \cite{wu2017influence}. }.

\section{Conclusion}
\label{sec:Conclusion}

We studied the robustness of flow networks consisting of $N$ lines against random attacks under a partial load redistribution model. In particular, when a line fails due to overloading, it is removed from the system and $(1-\varepsilon)$-fraction of the load it was carrying gets redistributed equally among all remaining lines while the remaining $\varepsilon$-fraction is assumed to be lost or absorbed. We derive recursive relations describing the dynamics of cascading failures for any attack size $p$, and identify the {\em final} fraction of surviving lines when the cascades stop. These findings are confirmed via extensive simulations. Among other things, we show that unlike the full redistribution case (i.e., $\varepsilon=0$), partial redistribution might lead to the order of transition at the critical attack size $p^{\star}$ changing from first to second-order, and a tricritical point emerges with respect to attack size $p$ and absorption/loss factor $\varepsilon$.

One of the most interesting findings of this paper is concerned with how system robustness can be {\em maximized} by properly choosing the distribution $p_{LS}$ that generates the initial load and free-space values of each line. We consider this problem when the mean load $\mathbb{E}[L]$ and mean free-space $\mathbb{E}[S]$ are fixed. First, we show that unlike the full redistribution case (i.e., when $\varepsilon=0$), the critical attack size $p^{\star}$ is not necessarily maximized by assigning every line the same free-space $\bE{S}$; depending on the fraction $\varepsilon$ of the load that is absorbed at each stage, we see that distributions other than Dirac-delta for $S$ may lead to higher critical points $p^{\star}$. Next, we consider the robustness metric proposed in \cite{scheider2011} that computes the {\em area} under the final system size $n_{\infty}(p, \varepsilon)$ over all possible attack sizes $0 \leq p \leq 1$; this amounts to computing the average response of the network to initial attacks of different size. We show that the system is most robust in the sense that the area metric is maximized, when the variation among the free-space of lines is minimized. In other words, the Dirac- delta distribution of free-space leads to the optimum robustness, irrespective of $\varepsilon$ and how load $L$ is distributed. Additionally, we test the robustness performance with respect to this area metric in finite flow networks with load data obtained from a real power network. We observe that allocating the free-space equally among the lines yields significantly higher performance when compared to well-known benchmarks despite the small size chosen for the network. Finally, we tested this robustness result in an extended topology-based redistribution model over an Erd\H{o}s-R\'enyi graph and we observed that uniformly allocating free-spaces promises to be optimal in this more general setting as well.      

There are many open problems one can consider for future work. For instance, the analysis can be extended to the case where the redistribution parameter $\varepsilon$ is not the same for all lines, but follows a given probability distribution. Similarly, $\varepsilon$ could be a time-varying parameter or it could depend on the extra load per line in the current stage. Such possibilities would allow us to obtain further understanding on the dynamical properties of the cascading failures and the mechanisms that could lead to a smooth failure. Additionally, it would be interesting to see if the robustness is still maximized with a Dirac delta type distribution on the free-space under various combinations of possibilities on $\varepsilon$. Variability in the redistribution as in \cite{hidalgo2002fracture} and stochastic behaviour in the loads as in \cite{kim2008fluctuation} also constitute interesting avenues for future research. In light of these directions, we will investigate relations among $\varepsilon$, the average number of neighbors and locality in redistribution for their effects on robustness against cascading failures. Finally, it would be interesting to study the partial redistribution model under {\em targeted} attacks rather than random failures.

\section*{Acknowledgments}
This research was supported in part by National Science Foundation grants CCF \#1422165 and CCF \# 1646526, by the Department of Energy grant DE-OE0000779, and by the Department of Electrical \& Computer Engineering at Carnegie Mellon University.

\end{document}